\documentclass{xray_symp}
\usepackage{graphics}
\usepackage{epsf}
\usepackage{psfig}

\begin{document}

\def\R{~ROSAT}
\def\RAS{\R all sky survey}
\title{Absorption and scattering properties in polarized bright soft X-ray 
selected ROSAT AGN}
\author{D. Grupe\inst{1} \and Beverley J. Wills \and D. Wills \inst{2}
\and Karen M. Leighly \inst{3}}  
\institute{Max--Planck--Institut f\"ur extraterrestrische Physik,
 Giessenbachstra{\ss}e, 85748 Garching, Germany
\and  Astronomy Department, University of Texas, Austin, TX 78712, U.S.A.
\and Columbia University, New York, NY 10027, U.S.A.}
\maketitle

\begin{abstract}

We have surveyed the optical linear polarimetric properties of 65 soft X-ray
selected ROSAT Active Galactic Nuclei (AGN).  
Most of these sources show low
polarization ($<$1\%) and no optical reddening. This is in 
agreement with the X-ray results suggesting a direct view on the
center. However, two of our AGN show polarization 
as high as 7--8\% and high
reddening. This optical reddening suggests a high column of dusty gas,
but why does this gas not absorb the soft X-rays? We suggest that the
dusty gas is actually ionized. There is evidence for these 'warm'
absorbers from X-ray absorption features seen in ROSAT and ASCA spectra.
\end{abstract}

\section{Introduction}

It is commonly accepted by most astronomers that all 
Active Galactic Nuclei (AGN) are powered by the
same type of engine: accretion of matter though an accretion disk onto a 
super-massive black hole. What kind of matter is it that fuels the engine
and how is it distributed? In order to try to answer this question we have to 
apply different observational techniques, such as the study of the X-ray 
properties and optical linear polarimetry. 

Our study of soft X-ray selected AGN (Grupe et al. 1998a,b) has shown that
AGN with steep X-ray spectra tend to have relatively narrow 
Broad Line Region (BLR) emission 
lines, weak Narrow Line Region (NLR)
emission lines, and strong FeII emission (see also the
article by Thomas Boller about Narrow-Line Seyfert 1
galaxies in these proceedings). 
Two hypotheses to explain the narrowness of their H$\beta$\ are  (i) that the 
systems are
seen at high inclination, so the highest velocity BLR is obscured by a dusty 
torus.
This is suggested by unified schemes for Seyfert 2s, in which a 
`hidden' Seyfert 1
nucleus is revealed by its scattered (polarized) spectrum (e.g. Antonucci 
\& Miller 1985,
Cimatti et al. 1993), and (ii) that the narrower H$\beta$\ is the result of 
a lower mass
black hole -- with the strong, rapidly variable soft-X-rays arising from an
unobscured view of the hotter inner regions of the accretion disk 
(Boller et al. 1996, Grupe et al. 1998a).  
The former
hypothesis predicts scattering polarization, while the latter predicts zero
polarization.
If this is so, we can expect to see very broad lines and continuum in
scattered, polarized light.  This was the original
motivation for our polarization survey.  On the other hand, 
our X-ray studies have shown that there is little or no absorption of 
soft X-rays by neutral elements in these AGN (Grupe et al. 1998a). 
This lack of absorption suggests a direct view to the center, which means
we would not expect these sources to be polarized.

\section{Sample and Observations}

The AGN sample we used for the X-ray and optical spectral studies (Grupe et al.
1998a, b) was based on complete identifications of all bright soft X-ray
sources from the ROSAT All-Sky Survey (RASS, Voges et al 1993, 1998, 
Thomas et al.  1998).  The RASS used the position sensitive
proportional counter (PSPC), sensitive between 0.2 and 2 keV.  Our selection
criteria were: PSPC count rate $\rm >~0.5~cts~s^{-1}$, hardness ratio 1 
$<$ 0.0, and $|b|>20^{\circ}$).
Adopting these criteria led to a sample of 110 AGN.
In addition to the PSPC X-ray spectra, our database includes medium-resolution
spectroscopy obtained at ESO, La Silla, Chile, and McDonald Observatory, Texas.
Details of data reduction, and the results are given by Grupe et al. (1998b).

For the polarimetry survey, we used a
broad-band polarimeter on the 2.1m telescope at McDonald Observatory.
More details about the observations can be found in Grupe et al. (1998c),
where we presented results for the sample defined above.    
We have complete polarization data for 65 objects. 
The sample is complete for the northern hemisphere plus
some objects of southern declination that are accessible from McDonald or 
from which information can be found in the literature. All in all,
two thirds were observed by us, and for 1/3 we took data from the literature
(mainly Berriman 1989, Berriman et al. 1990). 
For those objects that turned out to be highly polarized 
we also obtained spectropolarimetry at the 2.7m telescope at McDonald.

We compared the results for our soft X-ray AGN sample with those from
the optically 
selected Narrow-Line Seyfert 1 galaxies (NLSy1s) 
sample that Goodrich (1989) studied by spectropolarimetry.
For the Goodrich sample we also collected optical spectroscopic and ROSAT
X-ray data.  While his sample is biased towards optical properties, it is
unbiased with respect to soft X-ray properties, e.g. cold absorption.

\section{Survey Results}

\begin{figure}
\psfig{figure=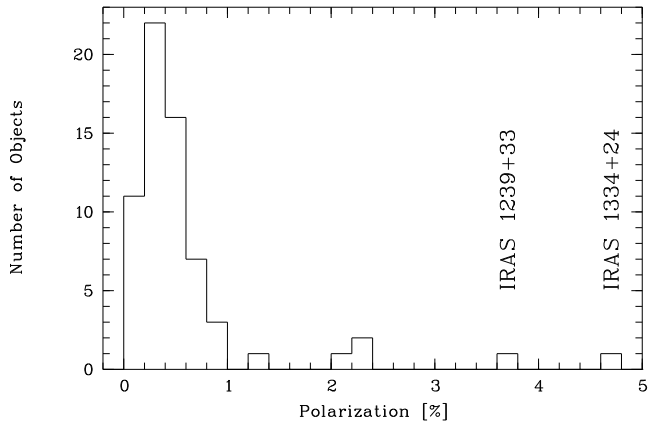,width=8.8cm,bbllx=10mm,bblly=10mm,bburx=280mm,bbury=195mm,clip=}
\caption{\label{distr} Distribution of the polarization of 65 objects 
of the soft X-ray AGN sample}
\end{figure}

Figure \ref{distr} displays the distribution of the degree of polarization
for the 65 objects of our soft X-ray selected sample.  Most of these
show no or only low degrees of polarization, p$<$1\%.  One object was 
found with p $\sim$1\% (CBS 126), three objects showed p$\sim$2\% (NGC\,1068,
H0439-272, Mkn\,766), and two sources with P$>$4\% were found
(IRAS\,F12397+3333, IRAS\,13349+2438).  While IRAS\,13349+2438 
has been extensively studied before (e.g. Wills et al. 1992, Brandt et al.
1996, 1997), IRAS F12397+3333 is new.

We were curious to find out
if there were any differences between polarized and
unpolarized AGN.  We did not find any differences in their
line widths or X-ray spectral indices compared with the rest of the sample.
However, the polarized AGN turned out to be the ones with high Balmer 
decrements and steep (reddened) optical continua.  These objects seem to be
affected by dust.

\subsection{The polarized AGN}
Six objects of the sample have shown polarization above 1\%: CBS 126,
NGC 1068, Mkn 766, H 0439-272, IRAS F12397+3333, and IRAS 13349+2438.
The degree of polarization of CBS 126 turned out to be wavelength independent.
Probably this can be explained by polarization through aligned dust grains and
not by scattering, which would show a stronger
wavelength dependence in the degree of
polarization. Scattering polarized AGN show a wavelength dependence because
we look for the dilution of the polarized light by unpolarized light. 
In the case where the unpolarized light absorbed by dust,
the unpolarized spectrum is reddened. Therefore,
the signature of scattering is an increase of the degree of 
polarization towards the blue.  
NGC 1068 is a special case in our sample, 
included because of its extended soft X-ray emission.
Mkn 766 is also one of the sources in Goodrich's NLSy1 sample.  We found
H 0439-272 to be polarized during our 1998 McDonald observations, so
detailed spectrophotometry is not yet available.
We concentrate here on the two `highly polarized' AGN
IRAS 13349+2438 and IRAS F12397+3333.  Both objects show a strong wavelength 
dependence in the degree of polarization (see Grupe et al. 1998c).
This is a sign of scattered light.  We observed both objects by 
spectropolarimetry.  While the 
results of this study have been already published for IRAS 13349+2438 (Wills
et al. 1992), we are preparing the results of the IRAS F12397+3333 data 
for publication now. 

The results of our spectropolarimetry
observation of IRAS F12397+3333 show that the there is s slightly higher degree
of polarization in the broad lines and a much lower degree of polarization
can be seen in the narrow lines. We discuss the significance of this later.

\begin{figure}
\psfig{figure=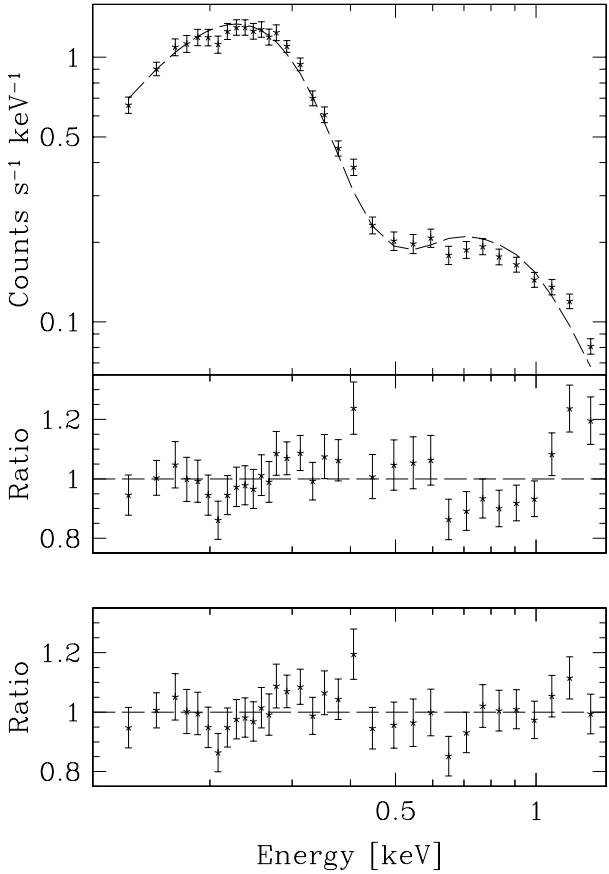,width=8.8cm,bbllx=1mm,bblly=1mm,bburx=200mm,bbury=280mm,clip=}
\caption{\label{rosat} ROSAT PSPC spectrum of the pointed observation of
IRAS F12397+3333. The upper panel shows a single power law fit to the data
($\alpha_{\rm X}$ = 2.0, $N_{\rm H}~=~2.8~10^{21}~cm^{-2}$, 
$\chi^{2}/\nu$ = 1.8). The fit can be significantly improved by 
adding an absorption edge of OVII at 0.74 keV (rest frame). We find an optical
depth of $\tau$ = 0.9$\pm$0.2, $\chi^{2}/\nu$ = 1.0. The ratio of this fit 
is given in the lower panel. 
}
\end{figure}

As noted before, the polarized AGN are those that are highly affected by 
dust reddening. The question is, should they not be affected by cold, neutral
absorption in soft X-rays as well?  Why do we see soft X-rays from these highly
reddened objects?  Objects such as IRAS 1509-21 from Goodrich (1989)'s
sample shows both high reddening and attenuation of soft X-rays by
neutral absorption. The answer may be that the absorption is by dusty,
ionized 'warm' gas (extra ionized absorbers in there).
This gas absorbs at intermediate X-ray energies ($\sim$ 0.5-3.0 keV), but is
transparent to soft X-rays.  If our objects are affected by warm absorbers we 
can expect to see their characteristic absorption features around 0.7-0.9 keV,
caused by highly ionized oxygen ions (O\,VII at 0.74 keV and O\,VIII at 
0.87 keV).
Brandt et al. (1996, 1997) have intensively studied the ROSAT and ASCA data
of IRAS 13349+2438 and found exactly those features in the X-ray spectra. 
Fortunately, IRAS F12397+3333 had been observed serendipitously, in a 
long ROSAT 
`pointed' exposure of a nearby source.
We found warm absorber features in the 
ROSAT spectrum of this source (see Figure \ref{rosat}).  In order
to get a more significant result we had 
ASCA observations of this source in 1998 June just a few days before 
this conference. From a quick look at these new data, we also find 
warm absorber features in the ASCA spectra.  
The details of the ROSAT and ASCA data as
well as the spectropolarimetry results will be presented in detail
(Grupe, Leighly \& Wills; Wills, Grupe, Leighly \& Wang in preparation).

\section{Discussion}
Most of the AGN we studied have shown polarization $<$ 1\%.  This favors a
direct, unobscured view to the nucleus. This is in agreement with our
finding of no cold absorption of soft X-rays and the flat optical, unreddened
spectra and the low Balmer decrements.  We can therefore rule out the 
hypothesis that the relatively narrow BLR lines found in NLSy1s are a result
of obscuration of the inner high velocity BLR clouds. 

However, the situation in the polarized AGN might be different.  All those 
AGN have shown high Balmer decrements and reddened optical spectra.  This 
suggests that they are affected by dust. Their polarization is wavelength
dependent. The explanation is that the scattered and therefore polarized 
light is diluted by a redder unpolarized component. The direct light
must go through some dust.  We can ask where the scatterers, warm absorbers,
and dust are located.
The answer to this question may be found by
spectropolarimetry.  We find that the BLR lines are only slightly more
polarized than the continuum, but the NLR lines are much less polarized.
Therefore we can conclude that the dust must be located somewhere between the
BLR and us, but it does not affect the NLR emission. It must be inside or just
outside of the BLR. 

The observation of central soft X-ray emission in these highly reddened AGN 
can be explained by the presence of warm, ionized absorbers.  We find those
features in the X-ray spectra of our polarized AGN. Leighly et al. (1997)
have even shown that scattering polarized AGN tend to show warm absorber
features in their ASCA data. On the other hand,
objects that do not show `warm' absorber 
features in their ASCA spectra, turn out to be low or unpolarized.
ASCA observations in connection with HST observations of objects 
that have shown a warm
absorber in their ASCA spectra even reveal UV absorption features 
-- suggesting 
ionized outflows from AGN (e.g. Mathur et al. 1997). Further studies
of these polarized soft X-ray selected AGN can help us to understand the 
structure, kinematics, and physical conditions on scales less than a few pc
in AGN.
With the new upcoming X-ray satellites a more detailed study of the 
absorption features will be possible to get more information about 
the kinematics of these absorbers. 

\begin{acknowledgement}
We would like to thank Prof. Dr. K. Beuermann (Uni-Sternwarte G\"ottingen)
for his support of this study. DG and BJW were supported by a grant 
from the Space Telescope Science Institute
(GO-06766) and NASA Long Term Space Astrophysics grant NAG5-3431.
This research has made use of the NASA/IPAC Extragalactic Database (NED)   
which is operated by the Jet Propulsion Laboratory, California Institute   
of Technology, under contract with the National Aeronautics and Space      
Administration.                                                            

\end{acknowledgement}

{}

\end{document}